\documentclass[conference, 9pt]{IEEEtran}
\IEEEoverridecommandlockouts

\usepackage{amsmath,amssymb,amsfonts}
\usepackage{algorithmic}
\usepackage{graphicx}
\usepackage{textcomp}
\usepackage{xcolor}
\usepackage{makecell}
\usepackage{booktabs}
\usepackage{multirow}
\usepackage[style=ieee,backend=biber,maxbibnames=99]{biblatex}
\def\BibTeX{{\rm B\kern-.05em{\sc i\kern-.025em b}\kern-.08em
    T\kern-.1667em\lower.7ex\hbox{E}\kern-.125emX}}

\AtEveryBibitem{%
  \clearfield{url}%
  \clearfield{urldate}%
}

\begin{document}

\title{SemanticBBV: A Semantic Signature for Cross-Program Knowledge Reuse in Microarchitecture Simulation}

\author{
\IEEEauthorblockN{Zhenguo Liu$^{1}$\textsuperscript{}, Chengao Shi$^{2}$, Chen Ding$^{3}$, and Jiang Xu$^{1}$\textsuperscript{\dag}}
\IEEEauthorblockA{$^{1}$Microelectronics Thrust, The Hong Kong University of Science and Technology (Guangzhou), China \\
$^{2}$Dept. of Electronic and Computer Engineering (ECE), Hong Kong University of Science and Technology, Hong Kong \\
$^{3}$Department of Computer Science, University of Rochester, United States}
\IEEEauthorblockA{zliu094@connect.hkust-gz.edu.cn, cshiai@connect.ust.hk, cding@cs.rochester.edu, jiang.xu@hkust-gz.edu.cn}
\thanks{\textsuperscript{\dag} Corresponding author.}
}

\maketitle

\begin{abstract}
For decades, sampling-based techniques have been the de facto standard for accelerating microarchitecture simulation, with the Basic Block Vector (BBV) serving as the cornerstone program representation. Yet, the BBV's fundamental limitations: order-dependent IDs that prevent cross-program knowledge reuse and a lack of semantic content predictive of hardware performance—have left a massive potential for optimization untapped.

To address these gaps, we introduce SemanticBBV, a novel, two-stage framework that generates robust, performance-aware signatures for cross-program simulation reuse. First, a lightweight RWKV-based semantic encoder transforms assembly basic blocks into rich Basic Block Embeddings (BBEs), capturing deep functional semantics. Second, an order-invariant Set Transformer aggregates these BBEs, weighted by execution frequency, into a final signature. Crucially, this stage is co-trained with a dual objective: a triplet loss for signature distinctiveness and a Cycles Per Instruction (CPI) regression task, directly imbuing the signature with performance sensitivity. Our evaluation demonstrates that SemanticBBV not only matches traditional BBVs in single-program accuracy but also enables unprecedented cross-program analysis. By simulating just 14 universal program points, we estimated the performance of ten SPEC CPU benchmarks with 86.3\% average accuracy, achieving a 7143$\times$ simulation speedup. Furthermore, the signature shows strong adaptability to new microarchitectures with minimal fine-tuning.

\end{abstract}


\section{Introduction}
Cycle-accurate simulation is a cornerstone of computer architecture research, but its notoriously slow speed---often orders of magnitude slower than native hardware---makes simulating realistic, trillion-instruction workloads prohibitively expensive~\cite{survey}. To overcome this bottleneck, sampling techniques have become widely adopted in both industry and academia. Among these, targeted sampling, exemplified by the seminal SimPoint methodology~\cite{simpoint} and its influential successors like BarrierPoint~\cite{BarrierPoint} and LoopPoint~\cite{looppoint}, has been particularly impactful. The core innovation shared across these works is the Basic Block Vector (BBV)~\cite{bbv}, a hardware-independent signature that captures program phase behavior by tracking the execution frequency of basic blocks.

\begin{figure}[htbp]
    \vspace{-2ex}
    \centering
    \includegraphics[width=\linewidth]{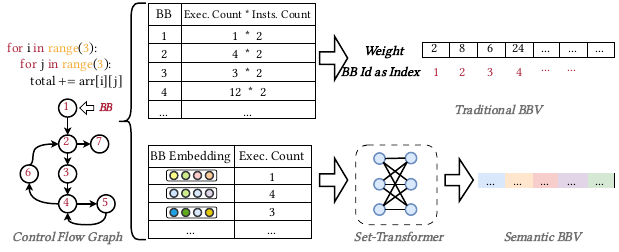}\vspace{-8pt}
    \caption{Comparison of Traditional BBV and SemanticBBV.}\vspace{-5pt}
    \label{fig:bbv_vs_semanticBBV}
\end{figure}

The classic BBV framework operates by first partitioning a program's execution into intervals. Within each interval, a signature is constructed as a high-dimensional vector. As illustrated in Figure~\ref{fig:bbv_vs_semanticBBV} (top), a nested for-loop is compiled into seven distinct basic blocks of assembly code. A basic block is a section of code with a single entry and exit. Each unique basic block encountered during execution is assigned a sequential ID, which serves as an index into this vector. The value at each index is then computed as the execution count of the corresponding basic block, often weighted by the number of assembly instructions it contains. Finally, these BBVs are grouped using clustering algorithms like K-means~\cite{kmeans}, with the cluster centroids selected as representative simulation points.

For two decades, the Basic Block Vector (BBV) has been the cornerstone of workload characterization. However, its fundamental design—assigning execution-order-dependent IDs—renders it inherently limited to single-program analysis. As acceleration techniques for individual programs approach their practical limits, the next major leap in simulation efficiency must come from reusing knowledge across different programs. This opportunity is vast: studies reveal that a small fraction of the SPEC CPU benchmark suite can often predict the performance of the entire set with high accuracy~\cite{spec2006_redundant, spec2017_redundant}. Yet, the BBV's single-program nature makes this immense potential for optimization unreachable, creating a critical bottleneck for next-generation simulation tools.

\begin{figure*}[!t]
    \vspace{-2ex}
    \centering
    \includegraphics[width=\textwidth]{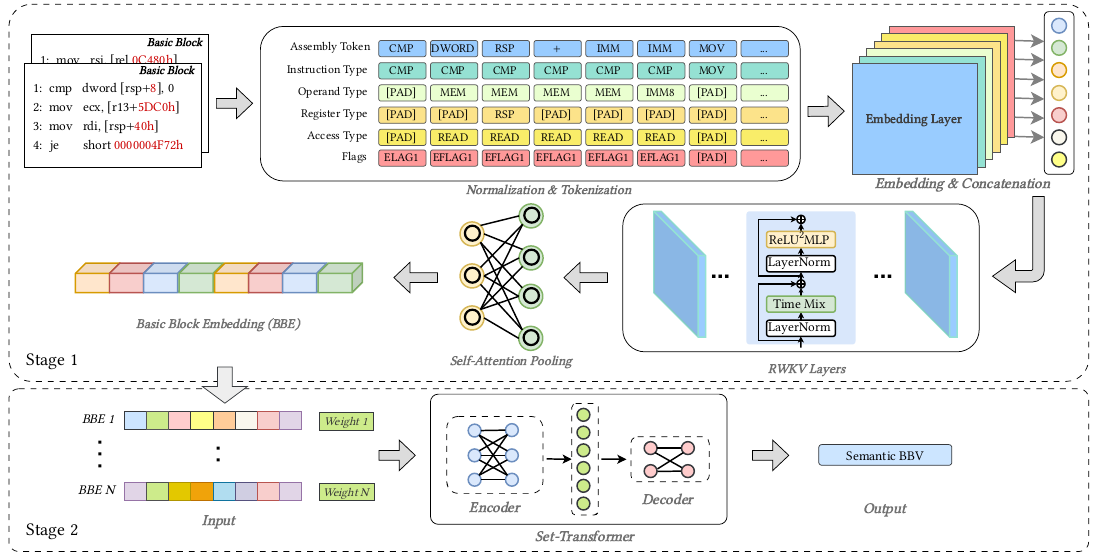}\vspace{-7pt}
    \caption{Overview of the SemanticBBV framework. The weighting mechanism introduced in Stage 2 is detailed in Figure \ref{fig:bbv_vs_semanticBBV}.}\vspace{-10pt}
    \label{fig:overview_of_semanticBBV}
\end{figure*}
To break this single-program barrier and unlock cross-program knowledge reuse, we propose SemanticBBV, a novel signature for cross-program reuse. Our key insight is to replace the fragile, order-dependent IDs with a representation based on the fundamental, hardware-agnostic semantic meaning of each basic block. This meaning, derived from assembly instructions, is inherently portable, allowing basic blocks to be compared regardless of their origin. As shown in Figure~\ref{fig:bbv_vs_semanticBBV} (bottom), we first learn rich embeddings for individual basic blocks and then aggregate them with an order-invariant Set Transformer~\cite{set-transformer}, producing a final, universally comparable signature.

We realize this vision through a two-stage neural framework. First, a lightweight RWKV-based~\cite{rwkv7} semantic encoder learns a dense embedding for each unique basic block. RWKV (Receptance Weighted Key-Value) is a recurrent architecture that achieves large language model–level performance while remaining parallelizable across sequence positions similarly to Transformers. Second, an order-invariant Set Transformer~\cite{set-transformer} aggregates these block embeddings into a compact signature. Critically, this stage is co-trained with a dual objective: maximizing signature distinctiveness while simultaneously performing Cycles Per Instruction (CPI) regression. This co-training ensures the final SemanticBBV is not only semantically aware but also highly predictive of hardware performance, making it a powerful tool for next-generation simulation acceleration.

Our design embodies a hybrid philosophy, motivated by the immense scale of dynamic execution traces. A purely end-to-end neural approach would be computationally prohibitive for intervals spanning millions of instructions. Therefore, SemanticBBV strategically combines the deep semantic understanding of neural networks (Stage 1) with the efficiency of statistical frequency counts (Stage 2), creating a solution that is both powerful and scalable for this challenging domain.

The contributions of this paper are summarized as follows:

\begin{itemize}
\item A novel, two-stage framework, SemanticBBV, that generates the first, to our knowledge, semantic and performance-aware signature designed explicitly for cross-program simulation reuse.
\item A lightweight yet powerful basic block encoder based on the RWKV architecture, which leverages a multi-dimensional tokenization scheme and novel pre-training tasks to achieve state-of-the-art performance on binary code similarity with significantly fewer parameters.
\item An order-invariant aggregation stage using a Set Transformer that is co-trained with a dual objective of signature similarity and CPI regression, directly embedding hardware performance sensitivity into the final signature.
\item A comprehensive evaluation demonstrating that SemanticBBV enables massive simulation speedups through cross-program knowledge reuse and can be efficiently adapted to new microarchitectures, validating its robustness and practical utility.
\end{itemize}
The remainder of this paper is organized as follows. Section II reviews related work in binary code similarity and learning-based sampling. Section III details our two-stage methodology for generating SemanticBBVs. Section IV presents a comprehensive evaluation of our framework, assessing its semantic quality, intra-program accuracy, cross-program estimation capabilities, and microarchitectural adaptability. Finally, Section V concludes the paper and discusses future directions.

\section{Related Works}
\subsection{Binary Code Similarity}\label{related_work_1}
Recent advances in binary code similarity have been heavily influenced by Transformer-based language models from NLP~\cite{bert}. Seminal works like PalmTree~\cite{PalmTree}, jTrans~\cite{jTrans}, kTrans~\cite{kTrans}, and UniASM~\cite{UniASM} have successfully adapted these architectures for assembly language. A key differentiator among these models is their tokenization strategy, which represents a trade-off between vocabulary size and contextual detail. 

At one extreme, PalmTree~\cite{PalmTree} employs a highly granular approach, decomposing an instruction into its most basic components (e.g., ``mov'', ``rdi'', ``,'', ``['', ``rsp'', ``]''). This yields a small vocabulary but results in long sequences. In contrast, models like jTrans and kTrans adopt a middle ground, partitioning instructions by opcode and operands, while UniASM~\cite{UniASM} treats each entire instruction as a single token, leading to a larger vocabulary but simpler contextual modeling.

\begin{sloppypar}
Beyond tokenization, these models employ various domain-specific pre-training objectives to capture rich code semantics for security-oriented tasks. These include Masked Language Modeling (MLM)~\cite{bert}, Context Window Prediction (CWP)~\cite{PalmTree} for learning code locality, and Def-Use Prediction (DUP)~\cite{PalmTree} for data flow.
\end{sloppypar}

However, these state-of-the-art models are ill-suited for accelerating microarchitecture simulation due to fundamental mismatches in scale and objective. First, their model architectures can only process short contexts (typically up to 512 tokens), rendering them impractical for directly processing simulation traces that can contain millions of instructions. More critically, their training objectives are designed to learn functional semantics, which do not necessarily correlate with hardware performance metrics like Cycles Per Instruction (CPI). This objective mismatch means that their embeddings may fail to distinguish program phases with different performance characteristics. Finally, the high computational cost of these large models conflicts with the efficiency demands of simulation acceleration.

\subsection{Learning-based Sampling Methods}
More recently, researchers have begun exploring deep learning to generate richer signatures for program sampling. A notable example is NPS (Neural Program Sampling)~\cite{nps}, which pioneers the use of a Graph Neural Network (GNN) on dynamic execution snapshots to create semantic-aware embeddings. NPS aims to overcome the context-insensitivity of traditional BBVs by modeling detailed code structures and runtime states.

However, the NPS framework introduces its own set of challenges. To represent long execution intervals (e.g., 10 million instructions), it first divides them into a vast number of micro-snapshots (each covering only 50 instructions) and then aggregates their embeddings through a two-stage process, which first employs mean-pooling to downsample the snapshots before an autoencoder generates the final embedding. This aggressive aggregation risks significant information loss, as critical temporal dynamics and transient behaviors within the larger interval are averaged out. Furthermore, while NPS targets improved sampling accuracy, its framework and evaluation remain focused on intra-program analysis. It does not explicitly address the challenge of creating a universal signature for cross-program knowledge reuse, which is a primary goal of our work. In contrast, our SemanticBBV is designed from the ground up to handle long, unordered sets of basic blocks without such lossy aggregation and is explicitly trained and evaluated for cross-program applicability.


\section{Methodology}
To overcome the limitations of traditional BBVs and enable cross-program knowledge reuse, we designed SemanticBBV, a two-stage framework that generates robust, performance-aware program signatures. As depicted in Figure~\ref{fig:overview_of_semanticBBV}, our approach first learns a dense, semantic representation for individual basic blocks (Stage 1), and then aggregates these representations into a final, order-invariant signature for an entire program phase (Stage 2). The following subsections detail the design and rationale behind each stage.

\subsection{Stage 1: Basic Block Semantic Embedding}
The goal of this stage is to transform a discrete, text-based basic block into a dense, semantic-rich vector—the Basic Block Embedding (BBE). This process, illustrated in the top panel of Figure~\ref{fig:overview_of_semanticBBV}, forms the semantic foundation of our entire framework.

We selected the basic block as our fundamental unit of analysis because it strikes an effective balance in granularity. It is coarse enough to make the total number of units manageable, yet simple enough to parse efficiently. Unlike complex functions, basic blocks lack nested structures and can be cleanly segmented from a binary stream by identifying jump instructions.

\subsubsection{Multi-dimensional Normalization and Tokenization}
A key challenge in binary analysis is representing assembly instructions in a way that is both expressive and computationally efficient. Prior work has faced a trade-off: fine-grained tokenization (like in PalmTree~\cite{PalmTree}) produces long, hard-to-learn sequences, while coarse, instruction-level tokenization (like in UniASM~\cite{UniASM}) struggles with large vocabularies (Section~\ref{related_work_1}). This coarse approach is also vulnerable to out-of-vocabulary (OOV) issues, where unseen instructions are mapped to a generic ``unknown'' token, leading to a critical loss of semantic information.

To address this, we designed a hybrid tokenization method that combines fine-grained granularity with rich, multi-dimensional features. Instead of relying on explicit boundary tokens (e.g., ``['', ``]'', ``,''), we implicitly encode structural information using six semantic dimensions: \textit{Assembly token}, \textit{instruction type}, \textit{operand type}, \textit{register type}, \textit{access type}, and \textit{flags}. The embeddings from each dimension are then concatenated, yielding a highly unique representation for each token that avoids ambiguity while keeping the vocabulary small. To further minimize vocabulary and prevent OOV issues, we also adopted a standard normalization strategy, replacing all immediate values and memory addresses with a generic \textit{IMM} token.

This combined approach not only reduces memory and computational demands but, more importantly, provides a richer feature space to express implicit semantics. For example, by tokenizing ``[rsp+IMM]'' as a single memory operand, our method captures its implicit dependency on the \textit{rsp} register—critical semantic information that is lost in the schemes of kTrans and UniASM. As shown in Table~\ref{tab:param_comparison}, this multi-dimensional strategy also results in a significantly smaller embedding layer than prior works.

\begin{table}[htbp]
    \centering
    \caption{Comparison of Embedding Layer Parameter Sizes}
    \label{tab:param_comparison}
    \begin{tabular}{c|c}
        \toprule
        \textbf{Model} & {\textbf{Parameters (M)}} \\ 
        \midrule
        kTrans          & 12.86 \\
        UniASM          & 10.75 \\
        jTrans          & 2.22  \\
        PalmTree        & 0.92  \\ 
        \textbf{Ours}   & \textbf{0.32}  \\
        \bottomrule
    \end{tabular}\vspace{-10pt}
\end{table}

\subsubsection{Backbone Architecture: RWKV}
To process the token sequences, we needed an architecture capable of efficiently handling the long and variable-length nature of assembly code. Traditional Transformers, with their quadratic complexity, are ill-suited for this task.

Therefore, we selected the RWKV (Receptance Weighted Key-Value) architecture~\cite{rwkv7} as our foundation model. As a linear Transformer, RWKV offers critical advantages for our domain: its \textbf{linear time complexity and constant memory usage} enable efficient training and inference. Its core mechanisms, such as the \textbf{time-mixing} for implicit positional handling and the \textbf{Delta Rule} for state updates, are ideal for capturing the state-dependent semantics of assembly code.

\subsubsection{Pre-training Task Design}
To enable our encoder to understand the semantics of assembly language, we first pre-trained it on a large corpus of unlabeled code. Since RWKV's unidirectional attention is incompatible with the popular Masked Language Model (MLM) task, we designed two novel, compatible self-supervised tasks:

\begin{itemize}
    \item \textbf{Next Token Prediction:} A standard autoregressive objective that forces the model to learn low-level syntax and patterns within a single instruction.
    \item \textbf{Next Instruction Prediction:} A novel task where the model, after processing one instruction, must predict the entire token sequence of the \textbf{next} instruction. This compels it to learn higher-level semantic flow between consecutive instructions.
\end{itemize}

\begin{figure}
    \centering
    \includegraphics[width=\linewidth]{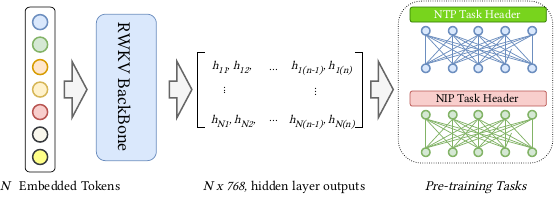}\vspace{-5pt}
    \caption{Pre-training Tasks}
    \label{fig:pre_training_tasks}
\end{figure}

As illustrated in Figure~\ref{fig:pre_training_tasks}, each task used a separate head (multi-layer perceptron) during pre-training to prevent the backbone from over-specializing. These heads were discarded before fine-tuning.

\subsubsection{Self-Attention Pooling and Basic Block Embedding Learning}

The encoder outputs a variable-length sequence of token embeddings for each basic block. To create a single, fixed-length BBE, a pooling mechanism is required. However, simple methods like mean-pooling are lossy; by averaging all token embeddings, they discard all information about token order and fail to weigh the relative importance of critical tokens like opcodes.

To address this, we employed a self-attention pooling layer~\cite{self_attention_pooling, self_attention_pooling2} that learns to weigh the importance of each token dynamically. Given the sequence of token hidden states $\mathbf{H} = [\mathbf{h}_1, \mathbf{h}_2, \dots, \mathbf{h}_N]$ from the encoder, we first compute an attention score $e_i$ for each token $\mathbf{h}_i$:
\begin{equation}
    e_i = \mathbf{u}_a^\top \tanh(\mathbf{W}_a \mathbf{h}_i^\top + \mathbf{b}_a)
\end{equation}

\noindent where $\mathbf{W}_a$ and $\mathbf{b}_a$ are the weights and bias of a fully connected layer, and $\mathbf{u}_a$ is a learnable context vector. These scores are then normalized into attention weights $\boldsymbol{\alpha} = softmax(\boldsymbol{e})$, ignoring any padding tokens. The final Basic Block Embedding $\mathbf{v}_{\text{BBE}}$ is the weighted sum of the token hidden states:
\begin{equation}
    \mathbf{v}_{\text{BBE}} = \sum_{i=1}^{L} \alpha_i \mathbf{h}_i
\end{equation}

Furthermore, a meaningful BBE must place semantically similar blocks closer in the embedding space than dissimilar ones. To enforce this property, we fine-tuned the entire encoder using a \textbf{Triplet Loss} objective~\cite{triplet_loss}. Following the methodology from~\cite{jTrans}, we constructed triplets where the anchor and positive samples were the same function compiled with different optimizations, and the negative sample was from a different function. This process ensures the resulting BBEs are robust to compiler variations and discriminative.

\subsubsection{Dataset}
For all semantic encoder development—including pre-training, fine-tuning, and evaluation—we exclusively used the large-scale BinaryCorp dataset~\cite{jTrans}. This dataset provides a rich and diverse source of assembly code, containing over 26 million functions compiled from more than 10,000 projects. These projects span a wide array of application types, including compilers, web browsers, and cryptographic libraries. Crucially, its functions are compiled with various optimization levels (O0, O1, O2, O3, Os), making it ideal for learning representations that are robust to compiler transformations. We strictly adhered to the official dataset splits, using the training split for all training phases and the held-out test split exclusively for the final evaluation to ensure an unbiased assessment of our model's generalization.

\subsection{Stage 2: Order-Invariant Signature Generation}
\subsubsection{Sequence or Set?}
A traditional BBV represents an \textbf{unordered set} of executed basic blocks and their frequencies, not a sequence. This set-like nature presents a critical challenge when using semantic embeddings: since there is no canonical ordering for the BBEs, a sequence-based model would produce different outputs for the same set of blocks, leading to meaningless similarity comparisons.

To solve this problem, we needed an architecture that is inherently \textbf{invariant to the input order}. This motivated our choice to move away from sequence-sensitive models like Transformers and RWKV, and instead select the \textbf{Set Transformer}~\cite{set-transformer}, which is specifically designed to operate on set-structured data.

\subsubsection{Set Transformer}

To meet the requirement for order-invariance, we leveraged the Set Transformer architecture~\cite{set-transformer}. Its encoder, composed of Self-Attention Blocks (SABs), processes all elements in the input set simultaneously, learning a context-aware representation for each BBE. Its decoder then aggregates these representations into a single, fixed-length signature using a Pooling by Multi-head Attention (PMA) block. Following the original paper's design, we constructed our encoder by stacking just two SABs and found this shallow architecture to be remarkably effective for our task.

\subsubsection{Training Tasks for Simulation-Aware Signatures}

A signature that is merely distinctive is insufficient for our goal; it must also be \textbf{sensitive to hardware performance}. A purely structural similarity (like that from BBVs) may not correlate with performance similarity (e.g., CPI).

To embed performance-awareness directly into the signature, we trained the Set Transformer with a multi-task objective that combines both structural and performance-based learning. This objective consists of three complementary loss terms:

\begin{itemize}
    \item \textbf{Triplet Loss:} This forms the foundation, ensuring that program intervals with similar execution patterns (based on traditional BBV similarity) are mapped closer in the new signature space. This preserves structural distinctiveness.
    
    \item \textbf{CPI Regression Loss:} To inject direct performance knowledge, we added a regression head to predict the ground-truth CPI of each interval, obtained from a Gem5~\cite{gem5, Lowe-Power:2020:gem5-20} simulation. We used the robust Huber Loss~\cite{Huber1992} for this task,  which is less sensitive to outliers than Mean Squared Error (MSE) and provides more robust training.

    \item \textbf{CPI Consistency Loss:} To ensure that signature similarity genuinely reflects performance similarity, we introduced this crucial regularization term. It penalizes pairs of signatures that are close in the embedding space but have a large difference in their actual CPI values. This forces the model to push apart structurally similar but performance-dissimilar intervals.
\end{itemize}

The final training objective is a weighted sum of these three components, allowing for flexible tuning of their relative importance:
\begin{equation}
    \mathcal{L}_{\text{total}} = \mathcal{L}_{\text{triplet}} + w_r \cdot \mathcal{L}_{\text{CPI\_Reg}} + w_c \cdot \mathcal{L}_{\text{consistency}}
\end{equation}
where $w_r$ and $w_c$ are the weights for the CPI regression and consistency losses. 

It is crucial to understand the conceptual implication of this training strategy. We can frame the relationship between code and performance as a function:
\begin{equation}
\label{eq:perf_model}
\text{Performance} = f(\text{Workload Behavior}, \textit{Hardware Context})
\end{equation}
In our framework, the SemanticBBV represents the ``Workload Behavior'', and the CPI regression task provides the ``Performance''. By training the model to map one to the other, we are implicitly forcing it to learn a representation of the function $f$ itself—that is, the characteristics of the underlying \textit{Hardware Context}.

Therefore, the goal is not to create a rigid, universal CPI predictor for one specific CPU. Rather, the primary purpose of this multi-task objective is to \textbf{inject a deep sensitivity to performance-relevant features} into the SemanticBBV. This makes our model and the signature a powerful and, most importantly, \textbf{adaptable} foundation. When targeting a new hardware environment, we do not need to retrain from scratch; we can simply fine-tune the model with a small amount of data from the new microarchitecture. This efficiently updates the learned mapping $f$, enabling rapid and accurate performance prediction across different hardware designs, as we will demonstrate in our evaluation (Section~\ref{Cross-Microarchitecture Adaptability}).

\section{Evaluation}

To comprehensively validate the SemanticBBV framework, our evaluation follows a deliberate, multi-stage progression. First, we assess the fundamental semantic quality of our Basic Block Embeddings, which form the basis of our signature (Section~\ref{Semantic Quality of Basic Block Embeddings}). Next, we demonstrate its efficacy in the traditional single-program context, establishing its parity with existing methods (Section~\ref{Intra-Program Simulation Accuracy}). We then showcase its primary contribution—the ability to facilitate cross-program analysis and knowledge reuse (Section~\ref{Cross-Program Simulation via Universal Clustering}). Finally, we test the model's versatility and generalization by evaluating its adaptability to new hardware environments (Section~\ref{Cross-Microarchitecture Adaptability}). Finally, we assess the computational efficiency of the framework to demonstrate its practical applicability (Section~\ref{Framework Performance}).

\subsection{Semantic Quality of Basic Block Embeddings}\label{Semantic Quality of Basic Block Embeddings}

The foundation of our framework is the ability of our Stage 1 encoder to produce high-quality Basic Block Embeddings (BBEs). To validate this, we evaluated our BBEs on the standard Binary Code Similarity Detection (BCSD) task using the test set from the BinaryCorp dataset~\cite{jTrans}. The task challenges a model to identify semantically identical functions that have been compiled with different optimization levels (e.g., O0 for no optimization and O3 for aggressive optimization). For each optimization pair being tested (e.g., O0/O3), we constructed 1,000 test cases. In each case, a function compiled at one level serves as the ``query'', and the model must find its counterpart, compiled at another level, within a large ``pool'' of distractor functions. For our baseline comparison, we utilized the officially released, pre-trained model weights for two established state-of-the-art (SOTA) models: UniASM~\cite{UniASM} and kTrans~\cite{kTrans}.

\begin{table}[htbp]
\centering
\caption{Average performance and model size on BCSD task. Best results are in bold.}\vspace{-5pt}
\label{tab:bcsd_summary_singlecol}
\begin{tabular}{@{}l r r cc@{}}
\toprule
\textbf{Model} & \textbf{Model Size} & \textbf{Pool Size} & \textbf{Avg. MRR} & \textbf{Avg. Recall@1} \\
\midrule
\multirow{2}{*}{UniASM} & \multirow{2}{*}{28M} 
& 100 & 0.566 & 0.488 \\
& & 10,000 & 0.314 & 0.274 \\
\midrule
\multirow{2}{*}{kTrans} & \multirow{2}{*}{110M} 
& 100 & 0.573 & 0.510 \\
& & 10,000 & 0.349 & 0.309 \\
\midrule
\multirow{2}{*}{\textbf{Ours}} & \multirow{2}{*}{\textbf{22M}} 
& 100 & \textbf{0.911} & \textbf{0.858} \\
& & 10,000 & \textbf{0.581} & \textbf{0.505} \\
\bottomrule
\end{tabular}\vspace{-10pt}
\end{table}

\begin{table}[htbp]
\centering
\caption{Detailed MRR results across optimization pairs, corresponding to the summary in Table~\ref{tab:bcsd_summary_singlecol}.}\vspace{-5pt}
\label{tab:bcsd_detail_singlecol}
\resizebox{\columnwidth}{!}{%
\begin{tabular}{@{}l r cccccc@{}}
\toprule
\textbf{Model} & \textbf{Pool Size} & \textbf{O0/O3} & \textbf{O1/O3} & \textbf{O2/O3} & \textbf{O0/Os} & \textbf{O1/Os} & \textbf{O2/Os} \\
\midrule
\multirow{2}{*}{UniASM} 
& 100 & 0.586 & 0.578 & 0.553 & 0.544 & 0.568 & 0.564 \\
& 10,000 & 0.334 & 0.319 & 0.302 & 0.304 & 0.301 & 0.322 \\
\midrule
\multirow{2}{*}{kTrans} 
& 100 & 0.596 & 0.592 & 0.579 & 0.582 & 0.561 & 0.529 \\
& 10,000 & 0.372 & 0.357 & 0.346 & 0.334 & 0.348 & 0.338 \\
\midrule
\multirow{2}{*}{\textbf{Ours}} 
& 100 & \textbf{0.906} & \textbf{0.906} & \textbf{0.913} & \textbf{0.902} & \textbf{0.917} & \textbf{0.919} \\
& 10,000 & \textbf{0.599} & \textbf{0.549} & \textbf{0.572} & \textbf{0.602} & \textbf{0.605} & \textbf{0.560} \\
\bottomrule
\end{tabular}\vspace{-5pt}
}
\end{table}

Performance is measured using two standard retrieval metrics: Recall@1 and Mean Reciprocal Rank (MRR). Recall@1 measures the percentage of queries for which the correct result is ranked first. MRR provides a more nuanced evaluation of the overall ranking quality. It is the average of the reciprocal of the rank of the first correct result for a set of queries ($N$):
\begin{equation}
\text{MRR} = \frac{1}{N} \sum_{i=1}^{N} \frac{1}{\text{rank}_i}
\end{equation}
where $\text{rank}_i$ is the position of the correct match for the $i$-th query.

As presented in Tables~\ref{tab:bcsd_summary_singlecol} and~\ref{tab:bcsd_detail_singlecol}, our encoder decisively outperforms the SOTA baselines. Our model achieves a significantly higher MRR and Recall@1 across both 100 and 10,000 candidate pools. Notably, our model's performance remains remarkably stable across all difficult optimization pairs (e.g., O0/O3), a task where competing models show noticeable degradation. This superior and consistent performance validates that our Stage 1 encoder produces high-quality semantic embeddings, forming a robust foundation for the final SemanticBBV signature.

\subsection{Intra-Program Simulation Accuracy}\label{Intra-Program Simulation Accuracy}
To validate SemanticBBV as a viable alternative to traditional BBVs, we first evaluated it in a standard single-program context. Our goal was to demonstrate \textbf{comparable} performance on the BBV's home turf before assessing its primary cross-program advantages.

\begin{figure}[htbp]
    \vspace{-2ex}
    \centering
    \includegraphics[width=\linewidth]{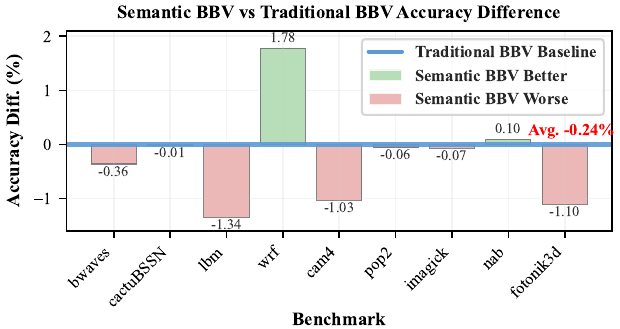}\vspace{-8pt}
    \caption{Comparison of simulation accuracy difference between SemanticBBV and traditional BBV on selected SPEC CPU 2017 floating-point benchmarks.}\vspace{-10pt}
    \label{fig:intra_program}
\end{figure}

Following the SimPoint~\cite{simpoint} methodology, we clustered 10-million-instruction intervals from the first 10 billion instructions of each SPEC CPU 2017 floating-point benchmark~\cite{spec2017} (excluding 654.roms due to execution errors). To ensure an unbiased evaluation, we specifically chose the floating-point suite, as our Stage 2 Set Transformer was trained on intra-program triplets derived exclusively from the integer benchmarks. Figure~\ref{fig:intra_program} compares the simulation accuracy of our SemanticBBV against the traditional BBV baseline.

Against the strong BBV baseline, which achieves \textbf{98.56\%} average accuracy (excluding the outlier 628.pop2), SemanticBBV's performance is highly comparable, with an average accuracy difference of just \textbf{-0.24} percentage points. Both methods struggle similarly on 628.pop2 ($\approx$63\% accuracy), suggesting a limitation of the K-means clustering for that specific workload rather than a flaw in either signature. This result confirms that SemanticBBV can serve as a drop-in replacement for traditional BBVs, justifying its evaluation in more complex cross-program scenarios.

\begin{figure}[htbp]
    \vspace{-2ex}
    \centering
    \includegraphics[width=\linewidth]{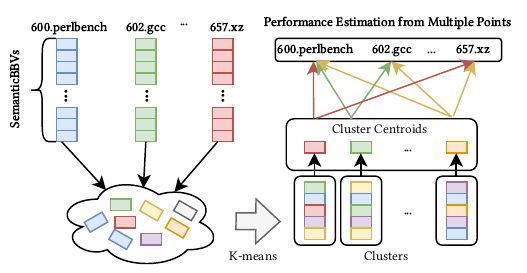}\vspace{-8pt}
    \caption{The workflow for cross-program performance estimation.}\vspace{-10pt}
    \label{fig:inter_program_semanticBBV}
\end{figure}

\begin{figure}[htbp]
    \vspace{-2ex}
    \centering
    \includegraphics[width=\linewidth]{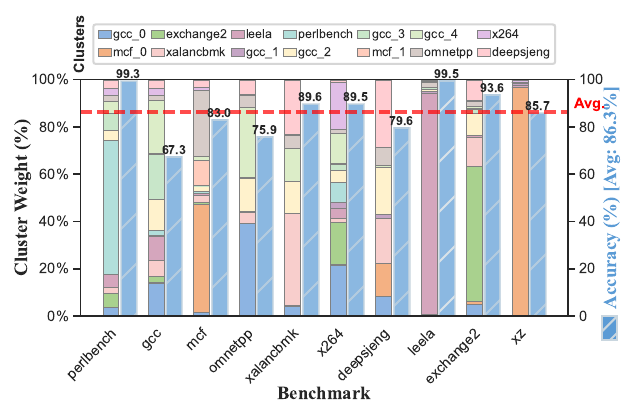}\vspace{-5pt}
    \caption{Behavior profiles of 10 SPEC CPU benchmarks across 14 universal clusters and CPI prediction accuracy using the cross-program estimation method.}
    \label{fig:cross_program_analysis}
\end{figure}
\subsection{Cross-Program Simulation via Universal Clustering}\label{Cross-Program Simulation via Universal Clustering}
The central promise of SemanticBBV is to enable knowledge reuse across programs, a capability absent in traditional BBVs. To demonstrate this, we conducted a cross-program estimation experiment on 10 SPEC CPU 2017 integer benchmarks~\cite{spec2017} (illustrated in Figure~\ref{fig:inter_program_semanticBBV}). We collected SemanticBBVs from one trillion instructions in total (100k intervals of 10M instructions) across all benchmarks and clustered this global pool into 14 universal behavioral archetypes using K-means, as suggested by SimPoint analysis~\cite{simpoint}. For each archetype, we then simulated only its single most representative interval to obtain a ground-truth CPI.

With this small set of 14 simulated points, we can estimate the performance of any of the 10 programs. We first characterize each benchmark by its ``behavioral fingerprint''—a vector representing its compositional breakdown across the 14 universal archetypes, as depicted in the left panel of Figure~\ref{fig:cross_program_analysis}. This fingerprint reveals a program's unique execution characteristics. For example, 602.gcc is highly heterogeneous, while 657.xz is remarkably uniform. Critically, although no interval from xz was chosen as a representative, its behavior is almost entirely (96.8\%) captured by a single cluster whose representative was sourced from 605.mcf. This powerfully illustrates cross-program semantic similarity.

Finally, a program's overall CPI is estimated via a weighted average of the 14 representative CPIs, using the program's behavior profile for weights. As shown in the right panel of Figure~\ref{fig:cross_program_analysis}, our method achieves 86.3\% average accuracy. By simulating just 140 million instructions (14 points $\times$ 10M), we successfully estimated the performance of one trillion total instructions—a 7143$\times$ speedup. This result validates that SemanticBBV creates a shared, semantic-aware space for effective knowledge transfer across disparate programs.
 
\subsection{Cross-Microarchitecture Adaptability}\label{Cross-Microarchitecture Adaptability}
A key measure of a truly powerful signature is its ability to generalize not just across programs, but also across different hardware. Our framework is designed for this adaptability. As established in our methodology, the Set Transformer is trained to generate a signature sensitive to the underlying hardware context. The key hypothesis is that this learned context can be efficiently updated for new architectures without retraining from scratch.

To validate this, we conducted a transfer learning experiment. Our initial model was trained using CPI data from Gem5's simple, in-order TimingSimpleCPU. We then targeted a significantly different microarchitecture: Gem5's complex, out-of-order O3 CPU. To adapt the signature generation process, we fine-tuned the Set Transformer and its CPI regression head using a remarkably small dataset—just 20\% of the initial 10B-instruction traces, randomly sampled from only two benchmarks (602.perlbench and 602.gcc). This fine-tuned model's performance was then evaluated on the first 10 billion instructions of the entire SPEC CPU 2017 integer suite.

\begin{figure}[htbp]
\vspace{-2ex}
\centering
\includegraphics[width=\linewidth]{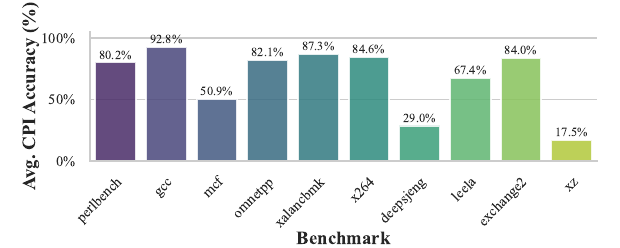}\vspace{-5pt}
\caption{CPI prediction accuracy on the SPEC CPU 2017 integer suite. The model was fine-tuned for a new microarchitecture (O3 CPU) using a small data subset from only two benchmarks (602.perlbench and 602.gcc).}\vspace{-7pt}
\label{fig:finetune_analysis}
\end{figure}

\begin{figure}[htbp]
\vspace{-2ex}
\centering
\includegraphics[width=\linewidth]{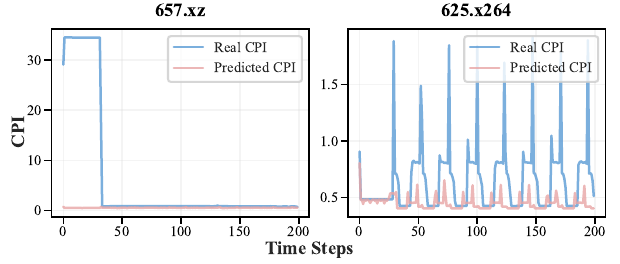}\vspace{-10pt}
\caption{Time-series comparison of real vs. predicted CPI for 657.xz and 625.x265 on the O3 CPU. The model captures the dynamic trend for x265 but misses the memory-driven spike in xz.}
\label{fig:cpi_prediction}
\end{figure}

The results, shown in Figure \ref{fig:finetune_analysis}, demonstrate powerful generalization. A compelling example is 625.x264. Despite having seen no data from this benchmark during fine-tuning, the model predicts its O3 CPI with 84.6\% accuracy. More importantly, as Figure \ref{fig:cpi_prediction} (right) illustrates, the model successfully captures the dynamic, periodic fluctuations in x264's performance profile. This proves that the fine-tuning process did not simply memorize patterns from the training programs; instead, it successfully updated the SemanticBBV's mapping to reflect the fundamental relationship between code semantics and performance on a complex out-of-order core.

Conversely, the experiment also cleanly highlights the current limitations tied to our training objective. The model performs poorly on memory-intensive benchmarks like 657.xz and 631.deepsjeng. Figure \ref{fig:cpi_prediction} (left) reveals why: our CPI-only model completely fails to predict the massive CPI spike (reaching over 30) caused by cold-start cache misses. Because the training objective did not include explicit memory-system metrics (e.g., L1/L2 miss rates), the resulting signature lacks sensitivity to these critical, memory-bound events.

This result is not a failure but an important insight that validates the framework's extensibility. For comprehensive design space exploration, the SemanticBBV can be enhanced by training it on a richer vector of hardware performance counters. This would create a signature sensitive not just to overall CPI but to the multifaceted interactions between code and specific microarchitectural features, such as the memory hierarchy and branch predictor, making it an even more powerful tool for architects.

\subsection{Framework Performance}\label{Framework Performance}
The SemanticBBV framework proves highly efficient, benchmarked on an NVIDIA RTX 4090 GPU. Its Stage 1 encoder processes tens of thousands of basic blocks per second, enabling the entire set of unique blocks from a large program like 602.gcc to be encoded in mere seconds. The Stage 2 aggregator is similarly fast, generating 2,000–3,000 signatures per second for 10M-instruction intervals. This throughput allows the one trillion instructions from our cross-program analysis to be processed in under a minute, confirming the pipeline is computationally lightweight and poses no bottleneck for large-scale analysis.

\section{Conclusion}
We have introduced SemanticBBV, a novel, two-stage framework that generates robust, performance-aware signatures for cross-program simulation reuse. By replacing the fragile, order-dependent IDs of traditional BBVs with deep semantic embeddings, SemanticBBV overcomes the single-program limitation that has constrained simulation acceleration for decades. Our hybrid approach, which combines a lightweight RWKV-based semantic encoder with an order-invariant Set Transformer, is designed to be both powerful and scalable to the immense size of dynamic execution traces.

Our evaluation confirms the framework's effectiveness. The final signature serves as a drop-in replacement for traditional BBVs in single-program analysis while, critically, enabling universal clustering and performance estimation across programs. This new capability yields massive efficiency gains: our cross-program estimation for ten benchmarks achieved 86.3\% average accuracy using just 14 representative points, resulting in a 7143$\times$ simulation speedup. Furthermore, the signature is adaptable, requiring only minimal fine-tuning to predict performance on new, complex microarchitectures.

SemanticBBV opens new avenues for efficient microarchitecture research. By enabling the reuse of simulation knowledge across entire benchmark suites, it paves the way for dramatically faster design space exploration and more intelligent, targeted workload analysis.

\printbibliography
\end{document}